\documentclass[sigconf]{acmart}

\usepackage{booktabs} 

\setcopyright{rightsretained}
\usepackage[ruled]{algorithm2e} 

\acmDOI{10.475/123_4}

\acmISBN{123-4567-24-567/08/06}

\acmConference[WOODSTOCK'97]{ACM Woodstock conference}{July 1997}{El
	Paso, Texas USA} 
\acmYear{1997}
\copyrightyear{2016}

\acmPrice{15.00}

\begin{document}
	
	\fancyhead{}
	\settopmatter{printacmref=false, printfolios=false}
	
	\title{Optimizing Sponsored Search Ranking Strategy by Deep Reinforcement Learning}
	
	\author{Li He}
	\affiliation{}
	\email{hl121322@alibaba-inc.com	}

	\author{Liang Wang}
	\affiliation{}
	\email{liangbo.wl@alibaba-inc.com}
	
	\author{Kaipeng Liu}
	\affiliation{}
	\email{zhiping.lkp@taobao.com}
	
	\author{Bo Wu}
	\affiliation{}
	\email{rhope@alibaba-inc.com}
	
	\author{Weinan Zhang}
	\affiliation{}
	\email{wnzhang@sjtu.edu.cn}
	
	

	\begin{abstract}
		Sponsored search is an indispensable business model and a major revenue contributor of almost all the search engines. From the advertisers' side, participating in ranking the
		search results by paying for the sponsored search advertisement to attract more awareness and purchase facilitates their commercial goal. From the users' side, presenting 
		personalized advertisement reflecting their propensity would make their online search experience more satisfactory. Sponsored search platforms rank the advertisements by a ranking
		function to determine the list of advertisements to show and the charging price for the advertisers. Hence, it is crucial to find a good ranking function which can simultaneously
		satisfy the platform, the users and the advertisers. Moreover, advertisements showing positions under different queries from different users may associate with advertisement
		candidates of different bid price distributions and click probability distributions, which requires the ranking functions to be optimized adaptively to the traffic characteristics. In this work,
		we proposed a generic framework to optimize the ranking functions by deep reinforcement learning methods. The framework is composed of two parts: an offline learning part which 
		initializes the ranking functions by learning from a simulated advertising environment, allowing adequate exploration of the ranking function parameter space without hurting
		the performance of the commercial platform. An online learning part which further optimizes the ranking functions by adapting to the online data distribution. Experimental results 
		on a large-scale sponsored search platform confirm the effectiveness of the proposed method.
	\end{abstract}

	\keywords{Sponsored search, ranking strategy, reinforcement learning}
	
	\maketitle
	
	\section{Introduction}
		
	Sponsored search is a multi-billion dollar business model which has been widely used in the industrial area \cite{Cheng10,Chakrabarti08}. In the commonly employed 
	pay-per-click model, advertisers are charged for users' clicks on their advertisements. The sponsored search platform ranks the advertisements by a ranking function and
	select the top ranked ones to present to the users. The price charged from the advertisers of these presented advertisements is computed by the generalized second price (GSP)
	auction mechanism \cite{GSP07,Myerson81} as the smallest price which is sufficient to maintain their allocated advertisement showing positions. Traditionally, the ranking 
	score of an advertisement is set to be its expected revenue to the sponsored search platform, computed as the product between the advertisers bid price and the predicted 
	click-through rate (CTR) of the user.
	
	The expected revenue based ranking function dominates the sponsored search area, and most of the existing methods focus on designing elaborate models to predict 
	the CTR \cite{Yu12,Li10,Cheng10,mcmahan2013ad}. In this work, instead of designing a CTR prediction model, we try an alternative way to add more flexibility in the ranking function for
	balancing the gain between users, advertisers and our advertisement platform. For the user experience, we recognize the engagement of users as their CTR on the 
	advertisements and a term is added to the ranking function which are monotonic of the users' CTR. To improve the advertisers' return on their spending, we add
	a term related to the users' expected purchase amount. The ranking function is then computed as the weighted sum of the two terms together with the expected revenue term.
	
	Although the ranking function can well represent the benefits of the platform, user and advertisers, it may not be directly related to the benefits of these players. First of all, owing to the 
	second price auction mechanism \cite{GSP07}, the advertisers are charged by the minimum amount of dollars they need to keep their advertisement position, not the amount of 
	money they bid. The price is determined by the competitiveness of the underlying advertisement candidates. Second, the CTR and CVR (i.e., conversion rate, the ratio of 
	buying behavior after each advertisement click) are predicted by a prediction model, which is generally biased and prone to noise because of being training on a biased distributed
	dataset \cite{Joachims17}. To link the real world benefits of the users, advertisers and platform directly to the ranking function, we propose a reinforcement learning framework
	to learn the ranking functions based on the observed gain in a `trail-and-error' manner. By treating the ranking function parameter tuning as a machine learning problem,
	we are able to deal with more complex problems including higher parameter space and traffic characteristic based tuning. Advertisement showing positions associated with 
	different queries and different users exhibit diverse characteristics in terms of distribution of bid price and CTR/CVR. Tuning the parameters according to traffic characteristics 
	would definitely improve the performance of the ranking function.
	
	The reinforcement learning \cite{Sutton98} is generally used in sequential decision making, which follows an explore and exploit strategy to optimize the control functions of the 
	agents. Starting from its proposal, it is mostly used in games and robotics applications \cite{Mnih15,Silver16,Li17} where the exploration can be done in a simulated or artificial environment. 
	When utilizing the algorithms in advertising platforms, we need to consider the cost for the exploration and try to perform exploration in a simulated environment. However, different from the games, simulating the advertisement serving and rewards (click, user purchase, etc.) is hard due to the large space of controlling factors like user intention, 
	traffic distribution changes, advertisers budget limitation etc. In this work, we build a simulated sponsored search environment by making the historical
	advertisement serving replayable. Specifically, the simulated environment is composed of an advertisement replay dataset (state in the terminology of reinforcement learning), 
	which for each advertisement showing chance, stores the full list of advertisement candidates together with their predicted CTR, CVR and the advertisers' bidding prices, and 
	a set of `virtual' exploration agents which simulate the advertisement results and users' response under different ranking functions (i.e., actions in reinforcement learning terminology).
	The reward for the exploration is computed by reward shaping methods \cite{Ng99}.
		
	However, there are two problems with the simulated environment: (1) the simulated state-action-reward tuple is temporally independent: it is hard to simulate the temporal correlation 
	between the response of a user on one advertisement and the response on the next presented advertisement; (2) the simulated response is inconsistent with online user response 
	to some extent. To solve problem (1), we employ an off-policy reinforcement learning architecture \cite{Timothy15} to optimize the advertisement ranking function, which does not  require the observation of the next states' action and reward. For problem (2), we use an offline calibration method to adjust the simulated 
	rewards according to the online observed reward. Moreover, after the offline reinforcement learning, we employ an online learning module to further tune the reinforcement learning model to better fit the real-time market dynamics.
	
	\noindent\textbf{Contributions.} In this work, we present our work of optimizing the advertisement ranking functions on a popular mobile sponsored search platform. The platform
	shows the search results to users in a streaming fashion, and the advertisements are plugged in fixed positions within the streaming contents. Our main contributions are summarized as follows:
	
	\begin{itemize}
		\item We introduce a new advertisement ranking strategy which involves more business factors like the user engagement and advertiser return, and a reinforcement learning framework to optimize the parameters of the new ranking function;
		\item We propose to initialize the ranking function by conducting reinforcement learning in a simulated sponsored search environment. In this way, the reinforcement learning can explore adequately without hurting the performance of the commercial platform;
		\item We further present an online learning module to optimize the ranking function adaptively to online data distribution.
	\end{itemize}

	\section{Related Works}
	
	Auction mechanisms have been widely used in Internet companies like Google, Yahoo \cite{GSP07} and Facebook \cite{Vickrey61} to allocate the advertisement showing positions 
	for the advertisers. The Internet advertising auction generally works in the following procedure: an advertiser submit a bid price stating their willingness to pay for an advertisement response 
	(click for a performance-based campaign, view for a branding campaign, etc.) The publisher ranks the advertisements according to their quality scores and bid prices, and presents the top
	ranked ones to the users together with the organic contents. The advertisers are then charged, for the response of the users, the minimum amount of dollars to keep the showing positions
	of their advertisements. The auction mechanism has been extensively studied in the literature both theoretically and empirically: Bejamin {\it{et al.}} in \cite{GSP07} investigate the 
	properties of generalized second price (GSP) and compare it with the VCG \cite{Vickrey61} mechanism in terms of the equilibrium behavior. 
	In \cite{Aranyak07}, the authors formulate the advertisement allocation problem as a matching problem with budget constraints, and provide theoretical proof that the 
	algorithm can achieve a higher competitive ratio than the greedy algorithm. To solve the ineffectiveness in {\it next-price} auction, 
	the authors in \cite{Gagan06} designs a truth-telling keyword auction mechanism. In \cite{Myerson81} followed by \cite{Benjamin10}, the reserved price problem is studied, including
	its welfare effects and its relation to equilibrium selection criteria. A field analysis on setting reserve prices in sponsored search platforms of Internet companies is presented 
	in \cite{Michael11}. Existing work generally focuses on the revenue effect of the auction mechanism to make it efficient in the bidding process and capable of profiting more revenue. In our 
	work, we use the generalized second price (GSP) auction for pricing. But since we are working on an industrial sponsored search platform, in consideration of long term return, 
	instead of maximizing the platform revenue only, we also add the user experience and advertiser utility terms into the ranking function.
	
	Reinforcement learning problem is basically modeled as a Markov Decision Process \cite{Sutton98}, which concerns with how agents adjust their policy to interact with the 
	environment so as to maximize certain cumulative rewards. Recently, with the combination of deep neural network \cite{lecun2015deep}, the reinforcement learning methods are able to
	work in the environment with high-dimensional observations by feeding
	large-scale experience data and training with powerful computational machines \cite{Li17}, and make breakthrough in many different areas including game of Go \cite{Silver16}, video games
	\cite{Wu17,Mnih16}, natural language processing systems \cite{Wu16,Su16} and robotics \cite{Schulman15,Levine16}. However, most of the existing applications 
	are conducted on simulated non-profitable platforms, where the experience data are easy to acquire and there is no restrict to try any agent policies and training schemes. 
	In a commercial system, however, the exploration of the reinforcement learning may bring in uncertainty in the platform's
	behavior, prone to loss of revenue, thus offline methods are a practical solution \cite{zhang2016collective}. In \cite{Li16}, Li and Lipton {\it et al.} design a user simulator under movie booking scenario. The simulator is designed on some rules and data observations.
	However, in our platform, there are far more factors (users, advertisers) to simulate. 
	For online advertising with reinforcement learning, the authors in \cite{amin2012budget} first propose to tune sponsored search keyword bid price in an MDP framework, where the state space is represented by the auction information, the advertisement campaign's remaining 
	budgets and life-time, while the actions refer to the bid price to set.
	Then in \cite{Zhang17}, the authors formulate the sequential bid decision making process in 
	the real-time bidding display advertising as a reinforcement learning problem. The method is based on assumptions that the winning rate depends only on the bid price and the actual clicks 
	can be well estimated by the predicted CTR. Hence, enabling the best bidding strategy can be computed in an offline fashion. In our work, we initialize the reinforcement learning model 
	using the offline simulated data, and combine the reward shaping method and online model update procedure to make the model consistent with the online data distribution.
	
	\begin{figure}[t]
		\includegraphics[width=0.5\textwidth]{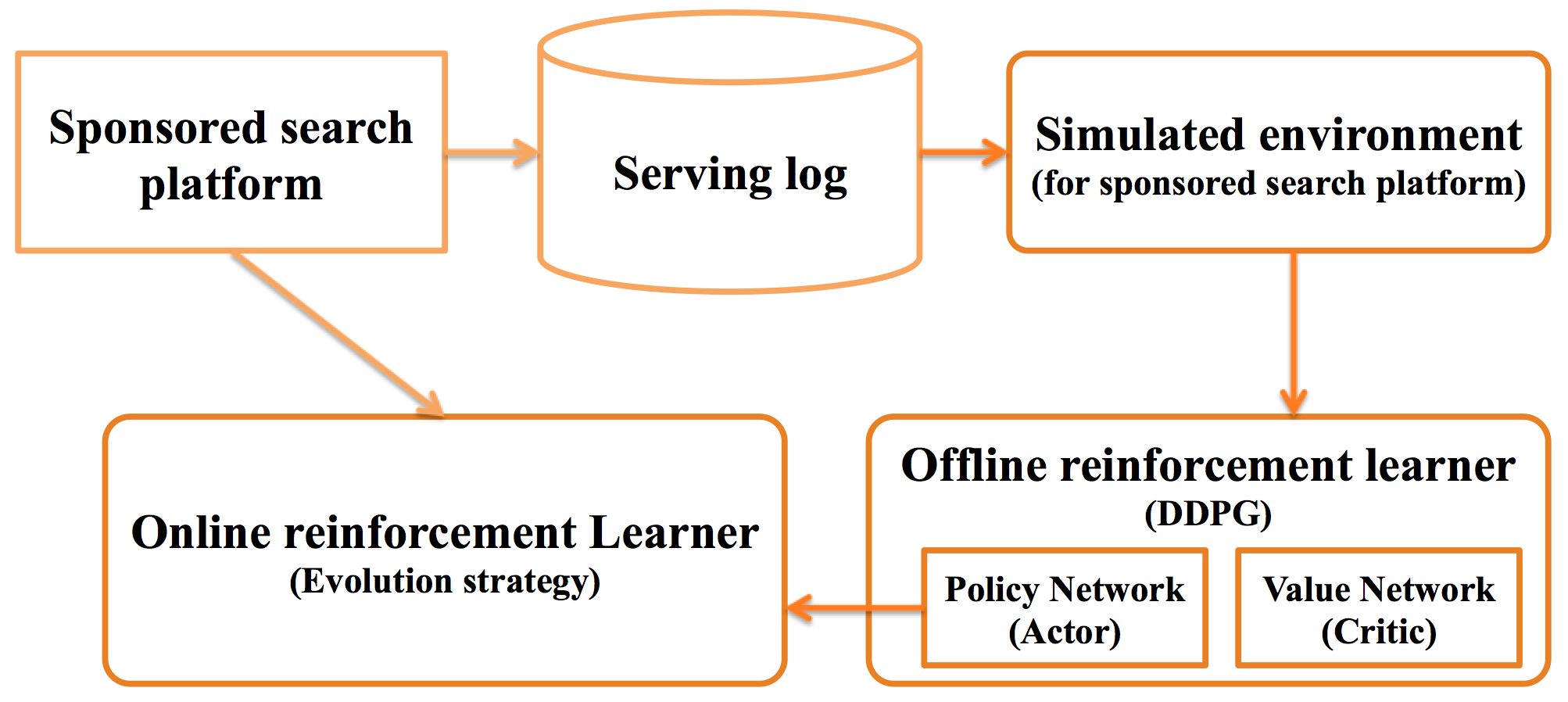}
		\caption{System flow chart.} \label{fig:flowchart}
	\end{figure}
	
	\section{System Overview}
	
	As shown in Fig.~\ref{fig:flowchart} (highlighted in orange), the whole system is composed of three modules: the offline sponsored search environment simulation module, 
	the offline reinforcement learning module and the online reinforcement learning module. The environment simulation module is used to 
	simulate the effect caused by changing the ranking function parameters, including re-ranking the advertisement candidates, showing the new top-ranked advertisement,
	and generating the users' response with respect to such changes. To allow adequate exploration, the offline reinforcement learning module collects training data by deploying
	randomly generated ranking functions on the simulated environment. An actor-critic deep reinforcement learner \cite{Timothy15} is then trained on top of these training data.
	To bridge the gap between the offline simulated data and the online user-advertiser-platform interaction, we build an online learning module to update the model by the
	online serving results.
	
	\subsection{Ranking Strategy Formulation}
	\label{sub:Ranking}
	
	We use the following ranking function to compute the rank score for advertisement $ad$
	\begin{equation} \label{eq:rankfunc}
	\phi(s, a, ad) = \underbrace{f_{a_1}(CTR) \cdot bid}_{\text{platform}} + a_2 \cdot \underbrace{f_{a_3}(CTR, CVR)}_{\text{user}} + a_4 \cdot \underbrace{f_{a_5}(CVR, price)}_{\text{advertiser}}
	\end{equation}
	where $s$ represents the search context including the search query, user demographic information and status of advertisement candidates for the advertisement showing chance.
	$bid$ and $price$ are the bid price and product price set by the advertiser for the advertisement $ad$ on the current query.
	$CTR$ and $CVR$ of $ad$ are predicted by the platform. $f_{a_i}~(i \in \{1, 3, 5\})$ performs nonlinear monotonic projection on 
	$CTR$ and $CVR$, and scalar $a_2, a_3$ are used to balance the weights between the three terms. Because our sponsored search platform charges the advertisers by 
	`click', the first term $f_{a_1}(CTR) \cdot bid$ can be seen as the expected revenue of the platform. We refer the users' preference to the presented advertisements as their 
	response ratio (CTR and CVR). Hence, the second term indicates the engagement of the users. The third term computes the expected return (expected user purchase amount) of the 
	advertiser by showing the advertisement, which measures the gain of the advertisers. According to the GSP rule, if the advertisement $ad$ is shown, its advertiser is charged
	\begin{equation} \label{eq:secondprice}
	click\_price = \frac {\phi(s, a, ad') - \big(a_2 \cdot f_{a_3}(CTR, CVR) + a_4 \cdot f_{a_5}(CVR, price)\big)}  {f_{a_1}(CTR)}
	\end{equation}
	where $ad'$ is the advertisement just ranked below $ad$. However, there is possibility that the numerator is a negative value. In our implementation, 
	we solve this problem by imposing a lower bound on the $click\_price$, just like the reserve price \cite{Michael11}, and since the platform revenue is always one of the optimization goal, the numerator is rarely negative
	in the experiments. Compared with the ranking function proposed in \cite{lahaie2011efficient}, which is generally used by Google and Yahoo, our ranking function~(\ref{eq:rankfunc}) 
	involves more parameters and commercial factors to consider, which can be used to optimize for more comprehensive commercial goals.
	
	The ranking function is parameterized by $a$ in consideration of the following issues: first, the predicted CTR and CVR are generally biased due to the imbalance training data,
	and need to be calibrated to make it consistent with the online user response; second, we charge the advertisers according to second price auction mechanism, which is lower
	than the bid price and computed according to the next ranked advertisement; third, the three terms may not be the same in numeric scale. The ranking function 
	optimization problem can be formulated as to predict the best parameter $a$ given the search context $s$ as
	\begin{equation} \label{eq:goal}
	\pi(s) = \arg\max_a R(\phi(s, a))
	\end{equation}
	where $R(\phi(s, a))$ is the reward given the ranking function $\phi(s, a)$. The reward can be defined as the sum of purchase amount, number of click and platform revenue 
	or any weighted combinations of the three terms during a certain period after the ranking function is operated, depending on the platform performance goal. Since in the reinforcement literature, $a$ is used to represent the action 
	of the learning agent, we take this notation to align with the literature.
	
	\subsection{Long-term Reward Oriented Ranking Function Optimization} \label{subsec:mdp}
	
	Reinforcement learning methods are designed to solve the sequential decision making problem, in order to maximize a cumulative reward. In the literature, it is generally 
	formulated as optimizing a Markov decision process (MDP) which is composed of: a state space $\mathcal{S}$, an action space $\mathcal{A}$, a reward space $\mathcal{R}$
	with a reward function $R: \mathcal{S} \times \mathcal{A} \rightarrow \mathcal{R}$ and a stationary transition dynamic distribution with conditional density $p(s_{t+1} | s_t, a_t)$ 
	which satisfies Markov property $p(s_{t+1} | s_1, a_1, ..., s_t, a_t) = p(s_{t+1} | s_t, a_t)$ for any state-action transition process $s_1, a_1, ..., s_t, a_t$. The action decision
	is made through a policy function $\pi_\theta: \mathcal{S} \rightarrow \mathcal{A}$ parameterized by $\theta$. The reinforcement learning agent interacts with the environment
	according to $\pi_\theta$ giving rise to a state-action transition process $s_1, a_1, ..., s_t, a_t$, and the goal of the learning agent is to find a $\pi_\theta^*$ which maximizes
	the expected discounted cumulative reward $r^\gamma = \sum_{{s_0} \sim p_0(s_0)}\sum_{k=0}^\infty \gamma^k \cdot$ $R(s_k, a_k)$ where $0 < \gamma < 1$ and $p_0(s_0)$
	represents the initial state distribution.
	
	
	The ranking function learning of a sponsored search platform has many special characteristics making it suitable to be formulated under the reinforcement learning framework. 
	First of all, during one user-search session, the sponsored search platform sequentially makes decisions on choosing ranking functions to presenting advertisements for the user. 
	Second, during the interaction with users, the platform collects users' response as rewards and balances between the exploration and exploitation to maximize the long term 
	cumulative rewards. Third, since the data distribution chances during the time, it requires online learning for adaptation. In this work, we choose the reinforcement learning 
	methodology to learn the ranking function in Eq.~(\ref{eq:goal}) for continuously improving the long term rewards. 
	Specifically, in our scenario, the state $s$ is defined as the search context of a user query including the query terms, query categories, user demographic information and online behavior.
	The sponsored search platform (i.e., reinforcement learning agent in our scenario) uses the ranking function as an action $a$ to rank the advertisements and interacts with a user to get the
	reward $r=R(s,a)$ as the combination of platform revenue, user engagement (quantified as user click, purchase, etc.) and advertiser's sale amount. Then the environment transits to the next state $s'$. The reinforcement 
	learning method optimizes the ranking function parameters through exploring and exploiting on the observed state-action-reward-state $(s,a,r,s')$ tuples.
	
	\section{Method}
	
	In the following sections, we will introduce the detail algorithms of the three modules in Fig.~\ref{fig:flowchart}.
	
	\subsection{Environment Simulation Module} \label{subsec:replay}
	
	On one hand, the performance of the reinforcement learning is guaranteed by adequate exploration; on the other hand, the exploration may bring uncertainty in the ranking
	functions' behavior and have performance cost. Since the algorithm is designed to run on a commercial sponsored search platform, we minimize the exploration cost by building
	a simulated sponsored search environment to training the reinforcement learning model in an offline manner.
	
	
	The sponsored search procedure is effected by many factors including the advertisers' budgets and bidding prices, the users' propensity and intention, making the procedure hard
	to simulate from scratch. In this work, instead of generating the whole sponsored search procedure, we propose to do the simulation by replaying the existing advertisement serving 
	processes and make them adjustable in terms of ranking function setting. 
	Given a user query, the platform proceeds by retrieving a big set of advertisements according to semantic matching criterion, predicting the CTR and CVR for these advertisements, and  
	computing the rank scores of them for advertisement selection and click price estimation. To make the advertisement ranking process replayable, 
	for each advertisement showing chance, we store the bidding information and predicted CTR, CVR of all the associated advertisement candidates. According to Eqs.(~\ref{eq:rankfunc})
	and (\ref{eq:secondprice}), the ranking orders and click prices for these advertisement candidates can be computed out of these replay information.
	The reward for showing an advertisement is simulated by the reward shaping approach \cite{Ng99} as the user response (like clicking 
	the advertisement or purchasing the product). For example, if our goal is to get more platform revenue and user clicks, the intermediate reward can be computed as
	\begin{equation} \label{eq:reward}
	r(s_t, a_t) = CTR \cdot click\_price + \delta \cdot CTR
	\end{equation}
	where $\delta$ is a manually tunable parameter to balance between the expectation of more user engagement (click behavior) or more platform revenue.
	To the best of our knowledge, most of the prediction algorithms concentrate on ordering the response rate of the advertisement instead of predicting the true value, and are trained
	on the biased data \cite{Joachims17}. As a result, there is a gap between the ground truth user response rates and the predict ones. To guarantee the reinforcement learning method 
	optimize towards the right direction, we minimize the gap by CTR and CVR calibration.

	\subsubsection{Reward Calibration} \label{subsub:reward}
	
	\begin{figure}
		\includegraphics[width=0.4\textwidth]{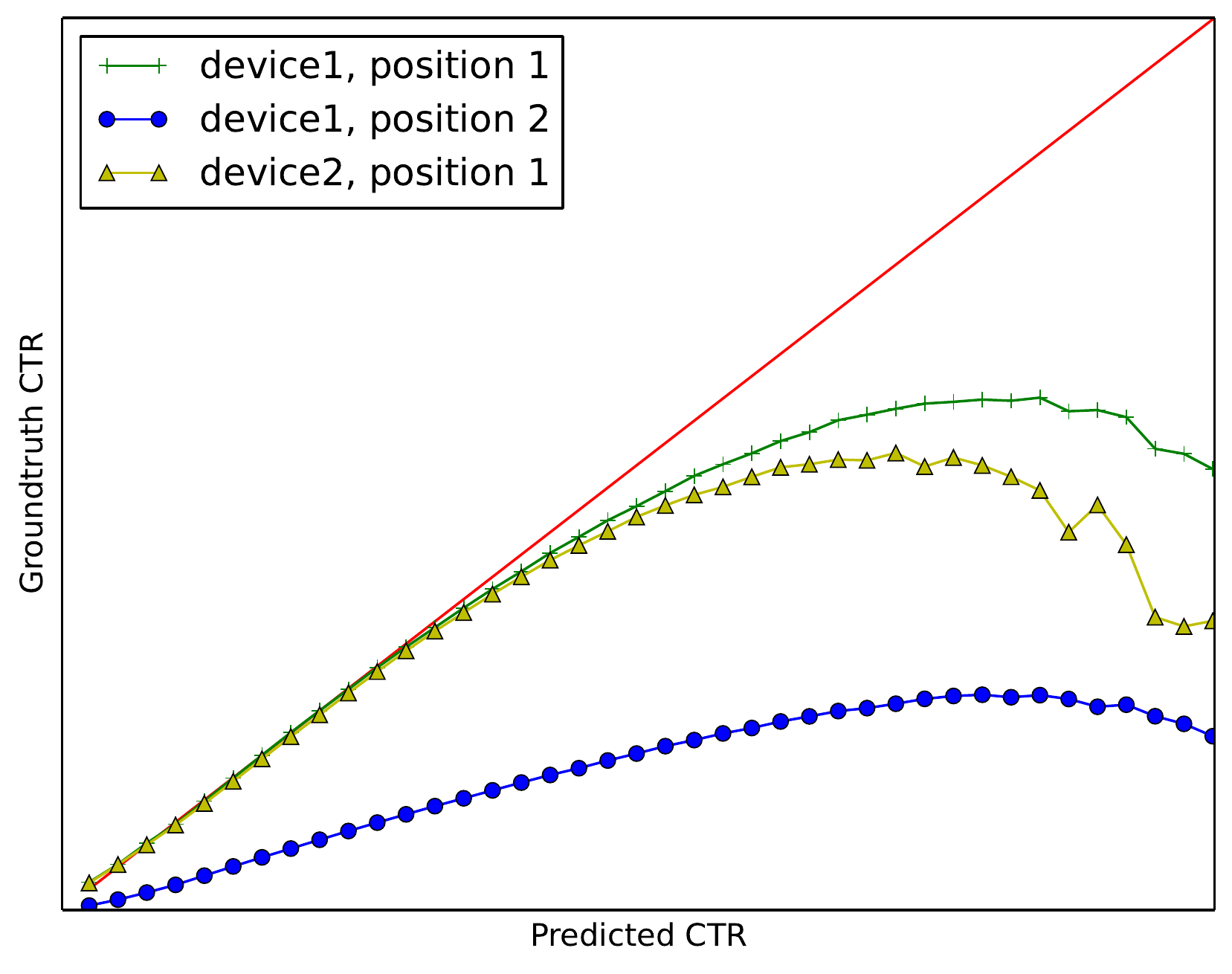}
		\caption{Illustration of the difference between the predicted CTR and the groundtruth value on two types of devices and two showing positions.} \label{fig:calibration}
	\end{figure}
	
	\begin{figure}
		\includegraphics[width=0.5\textwidth]{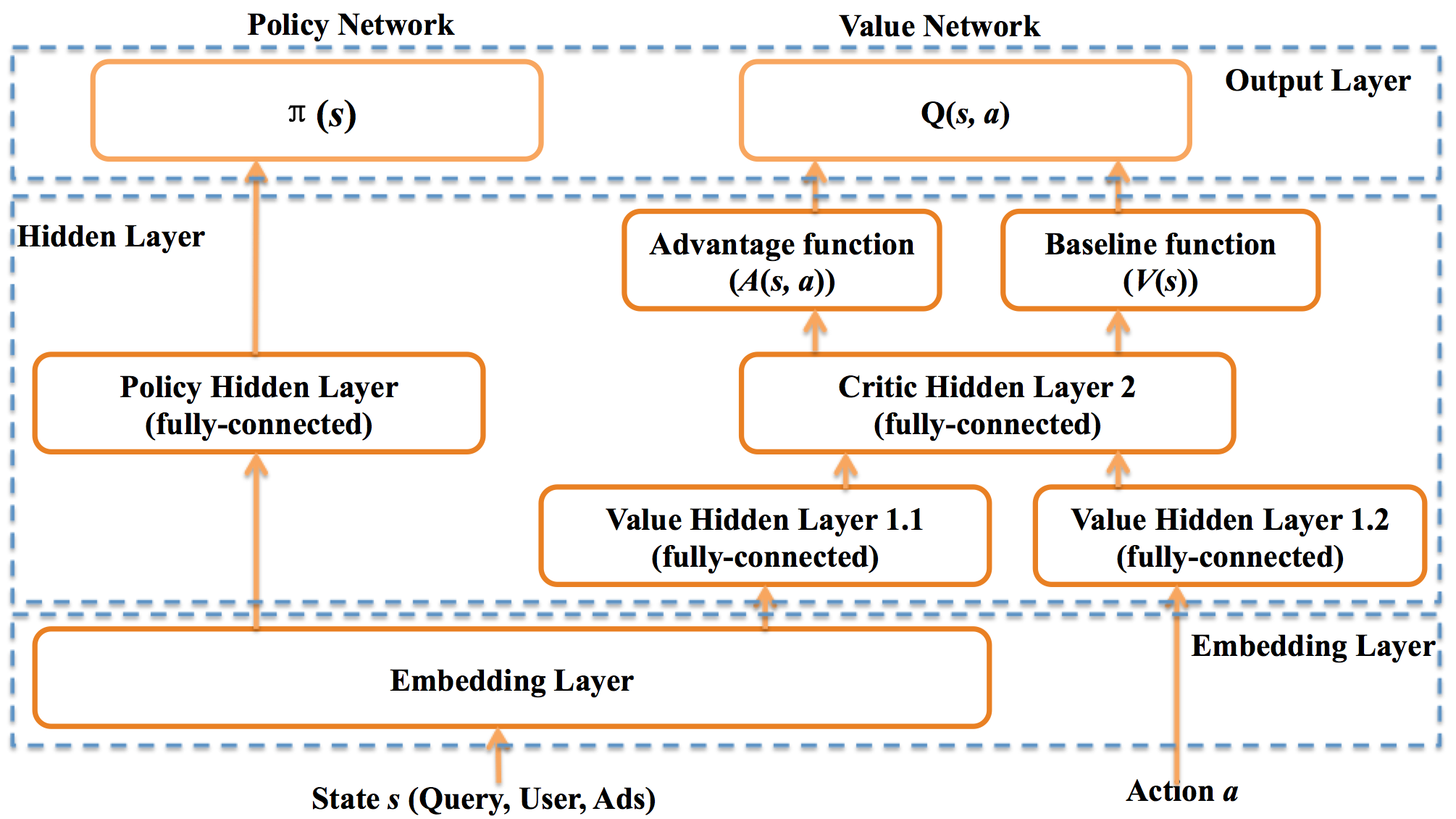}
		\caption{Illustration of the Actor and Critic network architectures used in our work.}
		\label{fig:DdpgNet}
	\end{figure}
	
	In Fig.~\ref{fig:calibration}, we illustrate the difference between the predicted CTR and groundtruth CTR on different device types and different advertisement showing positions.
	The advertisement showing chances (impressions) are grouped into bins by discretizing their predicted CTR values, and the groundtruth CTR of a bin is computed as the number of clicks achieved by the impressions in this bin divided by the number of impressions.
	It can be seen that, the predicted CTR and the groundtruth CTR exhibit diverse mapping relations on different context (e.g. device type, location). We manually select some 
	context features which effect the mapping relations most, and calibrate the predicted CTR as
	\begin{equation}
	\Gamma(CTR, \mathcal{F}) = \overline{CTR}
	\end{equation}
	where $\mathcal{F}$ refers to the manually selected feature fields like time and device type, $\overline{CTR}$ is the averaged ground truth CTR. To maintain the ordering relations of the predicted
	CTR on the set of advertisements, we employ the Isotonic regression method \cite{Best90} to compute the calibrated values for each $<CTR, \mathcal{F}>$ combinations. The method
	can automatically divide the CTR values into bins, targeting at minimizing the least square error between the predicted CTRs and the calibrated ones. By grouping the advertisement 
	showing results into bins according to the binned CTR values, the ground truth $\overline{CTR}$ for a bin is computed as the number of observed clicks in this bin divided by the number 
	of presented advertisements. CVR is calibrated in the same way.
	
	\subsection{Offline Reinforcement Learning for Ranking Function Initialization} \label{sub:offline}
	
	In this section, we will introduce the reinforcement learning problem formulation, the model architecture and the training method based on the simulated environment introduced above.
	Regarding to the MDP representation in Section~\ref{subsec:mdp}, for each advertisement showing chance, we define the state $s_t$ as the search context composed of three 
	types of features: (1) query related features like query ID, query category ID; (2) user demographic and behavior features including age, gender and aggregated click number on 
	certain advertisement; (3) advertisement related features, e.g. advertisement position. The action $a_t$ is the ranking function parameter vector $a$ in Eq.~(\ref{eq:rankfunc}), and 
	reward is defined as Eq.~(\ref{eq:reward}). Since user intention is difficult to model correctly, it is hard to predict whether a user would switch to another specific query after seeing 
	current advertisement. Thus we focus on the state transitions in each search session and omit the inter-session ones. To generate the next state $s_{t+1}$, we make simplification by assuming there is no change in query related features. The user behavior features are 
	updated by adding the expected behavior calculated out of the predicted CTR and CVR with calibration (refer to Section~\ref{subsub:reward}). For the advertisement related 
	features, we assume the user are continuously reading the streaming contents and advertisements one by one, and the advertisement related features are updated accordingly.
	
	Learning from the simulated environment introduced above gives rise to many special requirements for the learning algorithm. First of all, the simulation method above could only generate temporally
	independent state-action pairs, while lacking the capability of simulating the interaction between search sessions. This is because the state-action sequence of user-platform interaction is tangled. 
	The current user behavior is correlated with the previously presented advertisements and occurred user responses. The temporal independency of the training data requires the
	reinforcement learning method to support off-policy learning \cite{Thomas12}. Second, because the action space $\mathcal{A}$ is continuous  (refer to Eq.~(\ref{eq:goal})), it is practical to define 
	the deterministic policy function \cite{Silver14}. Moreover, the learning method should be capable of dealing with the complex mapping relation between action and rewards
	caused by the discontinuously distributed bid price. Taking these requirements into consideration, we use the Deep Deterministic Policy Gradient (DDPG) learning method in \cite{Timothy15}  
	as the learner to optimize our ranking function. The method supports off-policy learning, and combines the learning power of deep neural networks with the deterministic
	policy function property in Actor-Critic architecture.

	The DDPG algorithm iterates between the value network (critic network) learning step and the 
	policy network (actor network) learning step, the value network estimates the expectation of the discounted cumulative rewards from current time $t$ as (an example reward 
	function is shown in Eq.~(\ref{eq:reward}))
	\begin{equation} \label{eq:Q}
	Q_{\theta_Q}(s_t, a_t) = \mathbb{E}[r_t + \gamma \cdot r_{t+1} + \gamma^2 \cdot r_{t+2} + ... | s_t, a_t ]
	\end{equation}
	and the policy network calculates the best ranking function strategy given the current search context as
	\begin{equation} \label{eq:policy}
	a_t = \pi_{\theta_\pi}(s_t)~.
	\end{equation}
	The parameter $\theta_Q$ and $\theta_\pi$ refer to the weights and bias terms in the deep networks.

	\subsubsection{DDPG Network Architecture}
	
	The architectures of the value network and policy networks in our DDPG based reinforcement learning model is shown in Fig.~\ref{fig:DdpgNet}.
	Refer to Eqs.~(\ref{eq:Q}) and (\ref{eq:policy}), both the two networks have the current state $s_t$ as input. We represent all the features of $s_t$ (refer to Section~\ref{sub:offline})
	as ID features, and use a shared embedding layer to convert each of these ID features into a fixed-length dense vector and concatenate them to form the feature representation
	of $s_t$. The embedding vectors are initialized as random vectors and updated during the reinforcement learning process. For the policy network, we 
	connect the embedding layer to a fully-connected hidden layer with the Exponential Linear Units (ELU)~\cite{Clevert2015Fast} as activation function. In the experiment, we find 
	that when using activation functions like Sigmoid and ReLU, the output of nodes are easily move to numeric ranges with zero gradients, hindering the propagation of gradients
	along the network. By employing the ELU as activation function, the networks converge much faster. The hidden layer in the policy network is then concatenated to the output 
	layer by a sigmoid activation function. We also use clip method as \cite{Mnih15} to clip the output into a valid range to avoid over-learning.
	
	For the value network, its inputs are composed of the state features ($s_t$) and the action features (ranking function parameter vector $a_t$). Different from the state features,
	the action features are continuous. We connect them each to an independent fully-connected hidden layer with ELU as activation function. The two hidden
	layers are of the same number of nodes and are connected together to a higher level fully-connected hidden layer. In the output layer, we utilize a dueling network 
	architecture \cite{Wang2015Dueling} which
	divides the value function ($Q(s, a)$) in Eq.~(\ref{eq:Q}) into the sum of a state value function ($V(s)$) and a state-dependent action advantage function ($A(s, a)$),
	such that $Q(s, a) = V(s) + A(s, a)$. According to the insight of \cite{Wang2015Dueling}, the dueling architecture makes the learning of the value network efficient by identifying the 
	highly rewarded states and the states where the selected actions do not affect the rewards much. In our work, since our reward is defined on click prices (and product prices), it is 
	highly varied for different states and discontinuous with the change of advertisement candidates ordering. The variance in the rewards poses a big challenge for learning a stable policy function. 
	In our experiments (refer to Section~\ref{sub:OfflineSimExp} and Fig.~\ref{fig:OfflineExp1}), we observed that the dueling architecture makes the learning process converge
	more quickly. The observation coincides with the conclusion in \cite{Williams92}. Parameter setups of the network will be discussed in experiments.

	\subsubsection{Learning the Ranking Function from Simulated Environment} \label{sub:DdpgOffline}
	
\begin{figure}
		\includegraphics[width=0.5\textwidth]{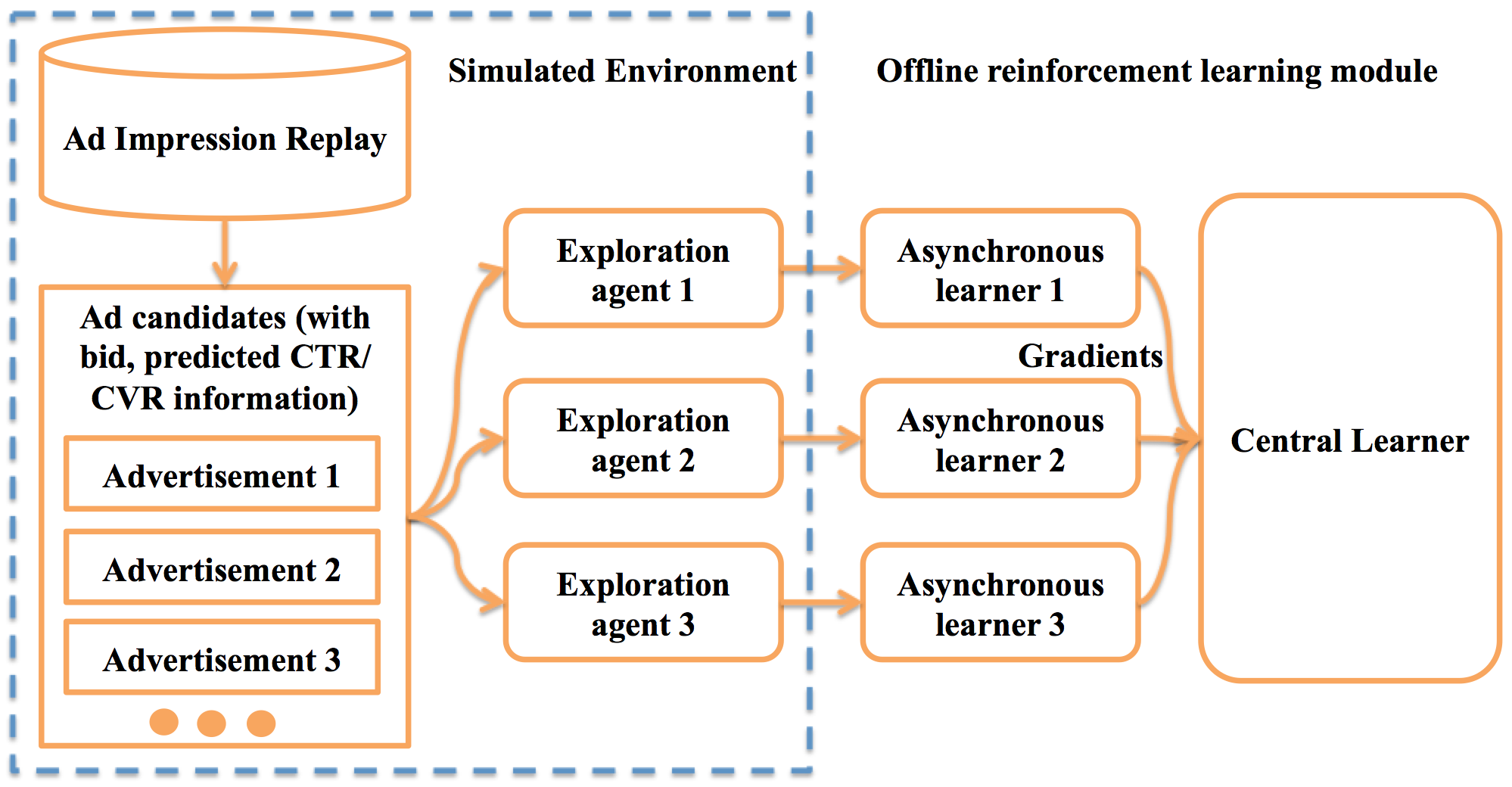}
		\caption{Illustration of the offline reinforcement learning framework.}
		\label{fig:offlinelearn}
	\end{figure}

	We employ the asynchronous training strategy \cite{Arun15} to train the DDPG model introduced above. As shown in Fig.~\ref{fig:offlinelearn}, the
	training data is sampled by multiple independent exploration agents. These training agents interact with the simulated sponsored search environment (Section~\ref{subsec:replay}) by sampling the advertisement serving replay data, trying different ranking functions (actions $a$'s) on the sampled replay data, and collecting
	the simulated rewards (Section~\ref{subsub:reward}) and state transitions (Section~\ref{sub:offline}) for these actions to build training tuples in the form 
	$\langle s_t, a_t, r_t, s_{t+1} \rangle$. The training tuples are then sent to local
	DDPG reinforcement learners to calculate the gradients of the value network parameters and policy network parameters and update these network parameters individually.
	The local learners also send their `local' gradients to update the `global' model parameters asynchronously. The algorithm is shown in Algorithm~\ref{alg:offlineAlg}. To 
	guarantee the independency and adequate exploration of the local reinforcement learners, we set the exploration agents to act by uniformly sampling from the ranking function parameter space.
	
	In our implementation, we use a distributed computing platform with around 100 GPU machines, and 5 central parameter servers to store the global model parameters.
	For the distributed computing platform, each process maintains an asynchronous DDPG learner, an exploration agent and a local model (copied from the global model). After several rounds of gradients computation from the simulated training data, the gradients are sent asynchronously to the central parameter servers, and the updated model
	parameters are sent back to update the local learners' copies.
	
	\begin{algorithm}[t]
	\SetAlgoNoLine
	\LinesNumbered
	\KwIn{Simulated transition tuple set $\mathcal{T}$ in the form $\psi=<s_t, a_t, r_t, s_{t+1}>$}
	\KwOut{Strategy Network $\pi_{\theta_\pi}(s_t)$}
	Initialize critic network $Q_{\theta_Q}(s_t, a_t)$ with parameter ${\theta_Q}$ and actor network $\pi_{\theta_\pi}(s_t)$ with parameter ${\theta_\pi}$\;
	Initialize target network $Q'$, $\pi'$ with weights $\theta_{Q'} \leftarrow \theta_{Q}$, $\theta_{\pi'} \leftarrow \theta_{\pi}$\;
	\Repeat{Convergence}{
		{Update network parameters ${\theta_Q}$, ${\theta_{Q'}}$, ${\theta_\pi}$ and ${\theta_{\pi'}}$ from parameter server}\;
		{Sampling subset $\Psi=\{\psi_1,\psi_2,...,\psi_m\}$ from $\mathcal{T}$}\;
		{For each $\psi_i$, calculate $Q^* = r_t + \gamma \cdot Q'(s_{t+1}, \pi'(s_t))$}\;
		{Calculate critic loss $L=\sum_{\psi_i \in \Psi}{\frac{1}{2} \cdot (Q^* - Q(s_t, a_t))^2}$}\;
		{Compute gradients of $Q$ with respect to ${\theta_Q}$ by $\bigtriangledown_{\theta_Q}Q = \frac{\partial L}{\partial \theta_Q}$}\;
		{Compute gradients of $\pi$ with respect to ${\theta_\pi}$ by $\bigtriangledown_{\theta_\pi}\pi  =\sum_{\psi_i \in \Psi}{ \frac{\partial Q(s_t, \pi(s_t))}{\partial \pi(s_t)} \cdot \frac{\partial \pi(s_t)}{\partial \theta_\pi} } =\sum_{\psi_i \in \Psi}{\frac{\partial A(s_t, \pi(s_t))}{\partial \pi(s_t)} \cdot \frac{\partial \pi(s_t)}{\partial \theta_\pi} }$}\;
		{Send gradients $\bigtriangledown_{\theta_Q}Q$ and $\bigtriangledown_{\theta_\pi}\pi$ to the parameter server}\;
		{Update ${\theta_Q}$ and ${\theta_\pi}$ with $\bigtriangledown_{\theta_Q}Q$ and $\bigtriangledown_{\theta_\pi}\pi$ for each global $N$ steps by gradients method}\;
		{Update ${\theta_{Q'}}$ and ${\theta_{\pi'}}$ by ${\theta_{Q'}} \leftarrow {\theta_{Q'}} + (1-\tau){\theta_Q}$, ${\theta_{\pi'}} \leftarrow {\theta_{\pi'}} + (1-\tau){\theta_\pi}$}\;
        }
	\caption{Asynchronous DDPG Learning}
	\label{alg:offlineAlg}
	\end{algorithm}

	\subsection{Online Reinforcement Learning for Ranking Function Updating} \label{Sec:OnlineLearning}
	
	\begin{figure}
		\includegraphics[width=0.45\textwidth]{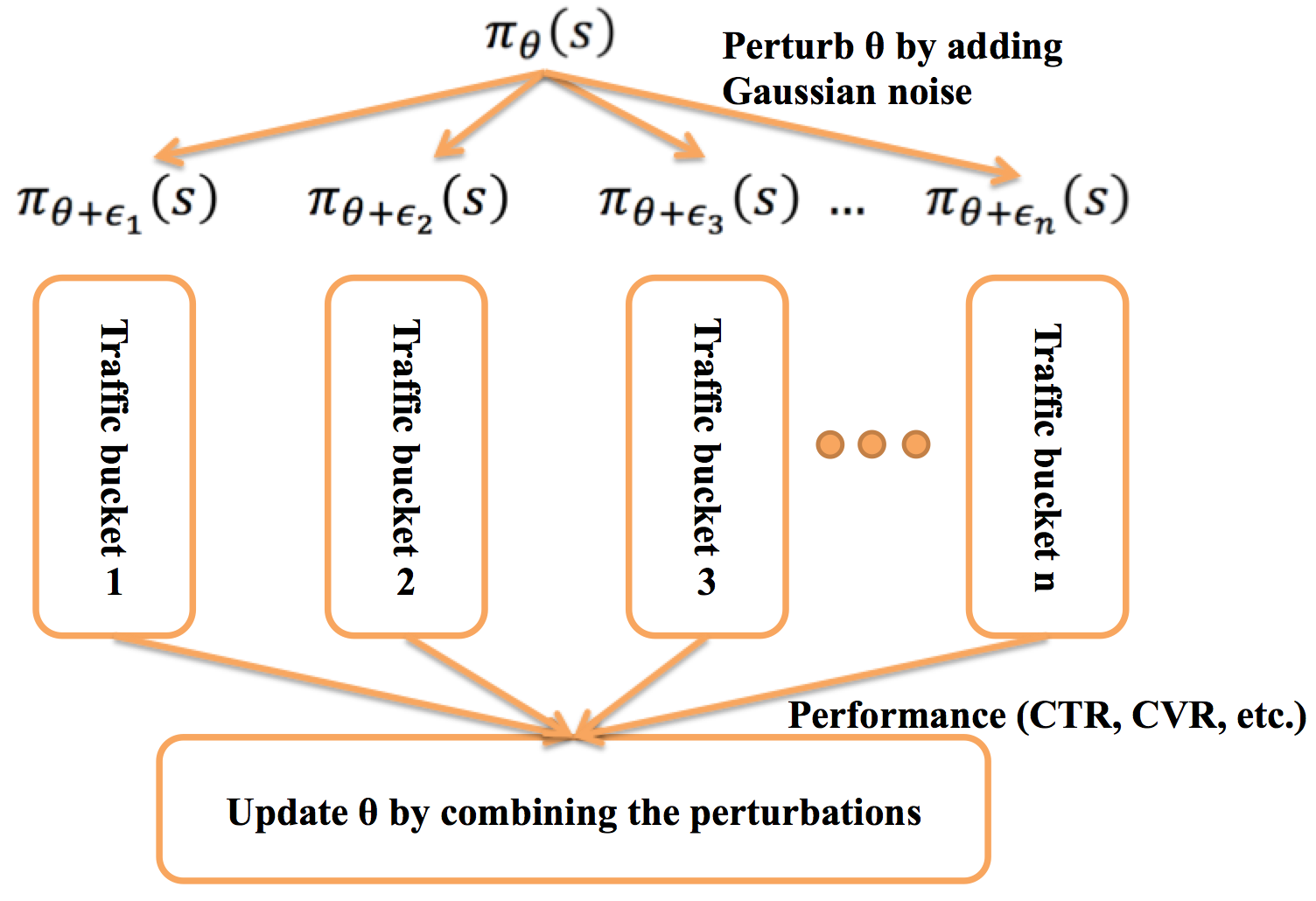}
		\caption{Illustration of the evolution strategy based online reinforcement learning method.}
		\label{fig:DdpgOnline}
	\end{figure}
	
	Despite the effort of reward calibration, the offline simulated environment is still inconsistent with the real online environment due to the dynamic data distribution and
	sequential correlation between the continuous user behavior. This inconsistency poses the online update requirement for the learned ranking function. However, directly using
	the asynchronous training framework in Section~\ref{sub:DdpgOffline} is not proper due to the speciality of the online updating: (1) the data distribution is different, where
	the online rewards are sparse and discrete (e.g. click or non-click, not the click expectation in simulated environment); (2) there are latency in reward collection, for 
	instance, a user clicks an advertisement immediately, but the purchase behavior may be postponed for several days \cite{chapelle2014modeling}.
	
	Regarding to these specialities, we introduce the evolution strategy \cite{salimans2017evolution} to update the parameters of the policy model. The evolution strategy based online
	updating method is illustrated in Fig.~\ref{fig:DdpgOnline}.
	We perform the following steps to online update the policy networks $\pi_{\theta_\pi}(s_t)$: (i) stochastically perturb
	the parameters $\theta_\pi$ by a Gaussian noise generator with zero mean and variance $\sigma^2$. Denote the set of $n$ perturbed parameters as 
	$\Theta_{\pi, \epsilon} = \{\theta_\pi+\epsilon_1, \theta_\pi+\epsilon_2, ..., \theta_\pi+\epsilon_n\}$. (ii) Hash the the online traffic into bins according to dimensions like user ID
	and IP address. For each parameter $\theta_{\pi, i} \in \Theta_{\pi, \epsilon}$, we deploy a policy network $\pi_{\theta_{\pi, i}}(s_t)$ on a traffic bin and get the reward according to 
	Eq.~(\ref{eq:reward}) as the weighted sum of platform revenue and the click number in this bin $R_i = total\_click\_price + \lambda \cdot click\_number$. However, in reality, the number of 
	advertisement showing number should not be exactly the same for each bin. We compute the relative value of the reward by dividing it with the number of served advertisements as $\overline{R}_i = \frac {R_i}  {served\_ad\_number}$.
	(iii) Update the parameter $\theta_\pi$ by the weighted sum of the perturbations as
	\begin{equation} \label{eq:onlineupdate}
	{\theta'}_\pi = \theta_\pi + \eta \frac{1} {n \sigma} \sum_{i=1}^{n} \overline{R}_i \epsilon_i
	\end{equation}
	where $\eta$ is the learning rate.
	It should be noted that regard to the online stability, we only use a small percentage of traffic (totally 2\% of the overall online traffic) for testing the performance of $\Theta_{\pi, \epsilon}$.
	
	
	The evolution strategy based method has several merits under our scenario. First of all, it is derivative-free. Since the rewards are discrete, it is hard to compute the gradient from the reward to the policy network parameters. Second, by fixing the seed of the random number generator, we just need to communicate the reward (a scalar) between the policy networks in local traffic bins and the central parameter servers. Thirdly, the method does not have intermediate reward requirement due to the homogeneity of these online traffic bins. Thus it can be deployed to optimize conversion related performance.

	\section{Experimental Results}
	
	We conduct experiments on a popular commercial mobile sponsored search platform which serves hundreds of millions of active users monthly covering a wide range of search categories. 
	To fully study the effectiveness of the proposed ranking function learning method, both analytical experiments on offline data and empirical experiments by deploying the learned ranking
	functions online are carried out. On the platform, the search results are presented in a streaming fashion, and the advertisements are allowed to be 
	shown on fixed positions within the streamed content. Since the search results are tangled with the advertisements, besides the platform revenue, one important issue we need 
	to deal with is the user experience. In the experiments, we design the immediate reward $r$ as
	\begin{equation} \label{eq:rewardexp}
	click\_price \cdot is\_click + \lambda \cdot is\_click
	\end{equation}
	where $click\_price$ is the amount we charge the advertisers according to generalized second price auction, and $is\_click$ is a binary number indicating whether the advertisement is
	clicked (1) or not (0). $\lambda$ is manually set according to the average $click\_price$ to balance between the platform revenue goal and the user experience goal. We can also add 
	the advertisers' satisfactory term like purchase price. But because we test our model on a small percentage of traffic online (2\% traffic), the purchase amount is highly varied according 
	to our observation. In the current experiments, we do not show the purchase optimized results and leave it as a future work when we ramp up our test traffic amount.
	
	
	\subsection{Experiments on Offline Data}
	\label{sub:OfflineSimExp}
	
\begin{figure*}
		\includegraphics[width=0.9\textwidth]{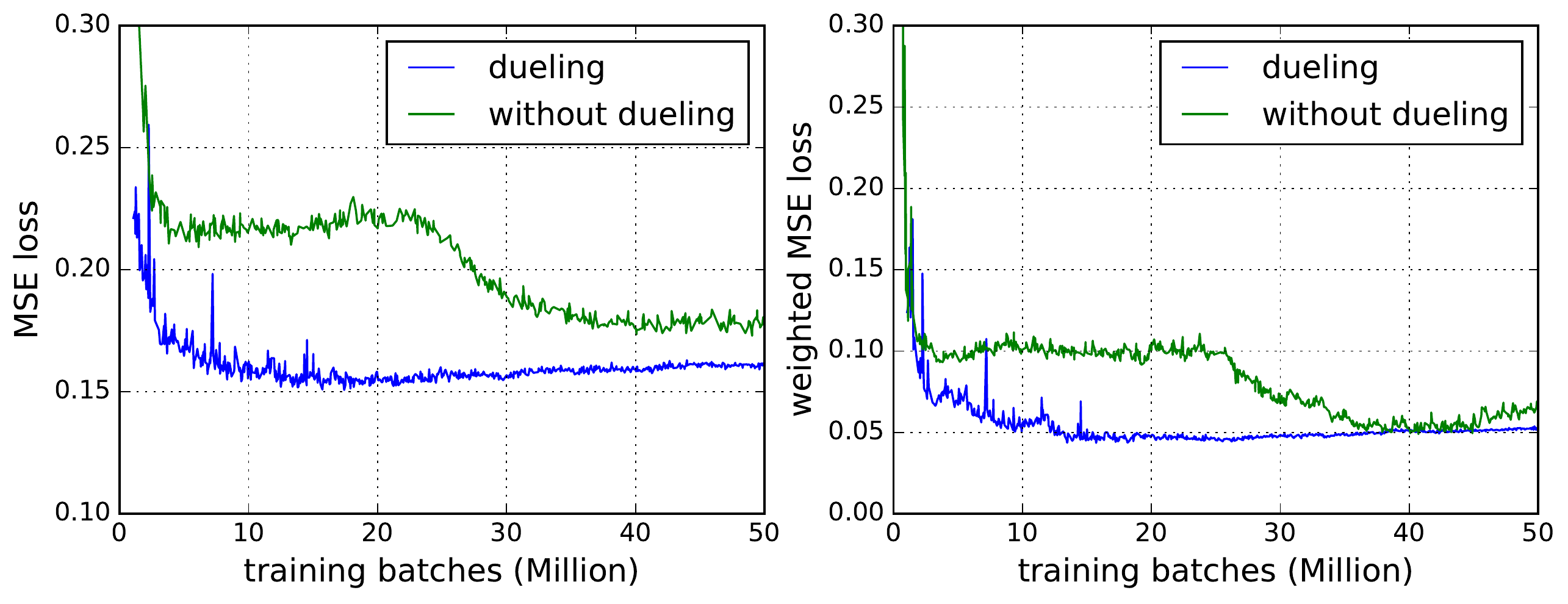}
		\caption{Comparison of the convergence speed of training by utilizing different network architectures. (1) Averaged squared error difference between strategy parameters of DDPG and the searched results; (2) advertisement impression weighted squared error difference between strategy parameters of DDPG and the searched results. `dueling' is the trained result using the dueling network structure and `without dueling' is the result without using the dueling network structure.}
		\label{fig:OfflineExp1}
	\end{figure*}

	
	Since we are learning on a biased sampled data, the mapping relation between the reward and ranking function parameters is complex due to the highly varied distribution of bidding price. The convergence property of the proposed method is worthy of studying. In the offline experiments, we study the convergence property of the proposed method and the effect of using different architectures and different super parameters on the speed of convergence. We employ an analytical method to verify whether the proposed method can converge to the `right' ranking function. In the experiment, a simple state representation (only query + advertisement position) is utilized such that from the simulated data, it is computational feasible to perform brute force search to find the best parameters of the ranking function (refer to Eq.~(\ref{eq:rankfunc})). The brute force method proceeds by uniformly sampling the parameters in $\theta$ at a fixed step size for each replay sample, computing the rewards (refer Eq.~(\ref{eq:rewardexp})) based on the method in Section~\ref{sub:DdpgOffline}, and finding the best parameter $\theta^*$ from these samples according to the aggregated rewards. For training the reinforcement learning model, we encode both the query IDs and advertisement position IDs into 8-dimension embedding vectors. As a result, our embedding layer consists of a 16-dimension feature vector. For the critic hidden layer, we utilize two full-connection units, each of which has 500 nodes, and ELU~\cite{Clevert2015Fast} as the activation function. We use the same settings for the actor hidden layer except using 100 nodes in each layer. The learning rate for network parameter, target network parameter and regularization loss penalty factor are set to be 1.0e-5, 0.01 and 1.0e-5 respectively. The $\lambda$ is set to be the average of the $click\_price$ calculated out of the data log. Experimental results are presented in Fig.~\ref{fig:OfflineExp1}. The performance of the proposed method is measured by the squared error between the learnt ranking function parameters and the `best' parameters found by brute force method. From the results, we can see, the proposed method could converge gradually to the best ranking function as the training process goes on.
	
\begin{figure}
		\includegraphics[width=0.5\textwidth]{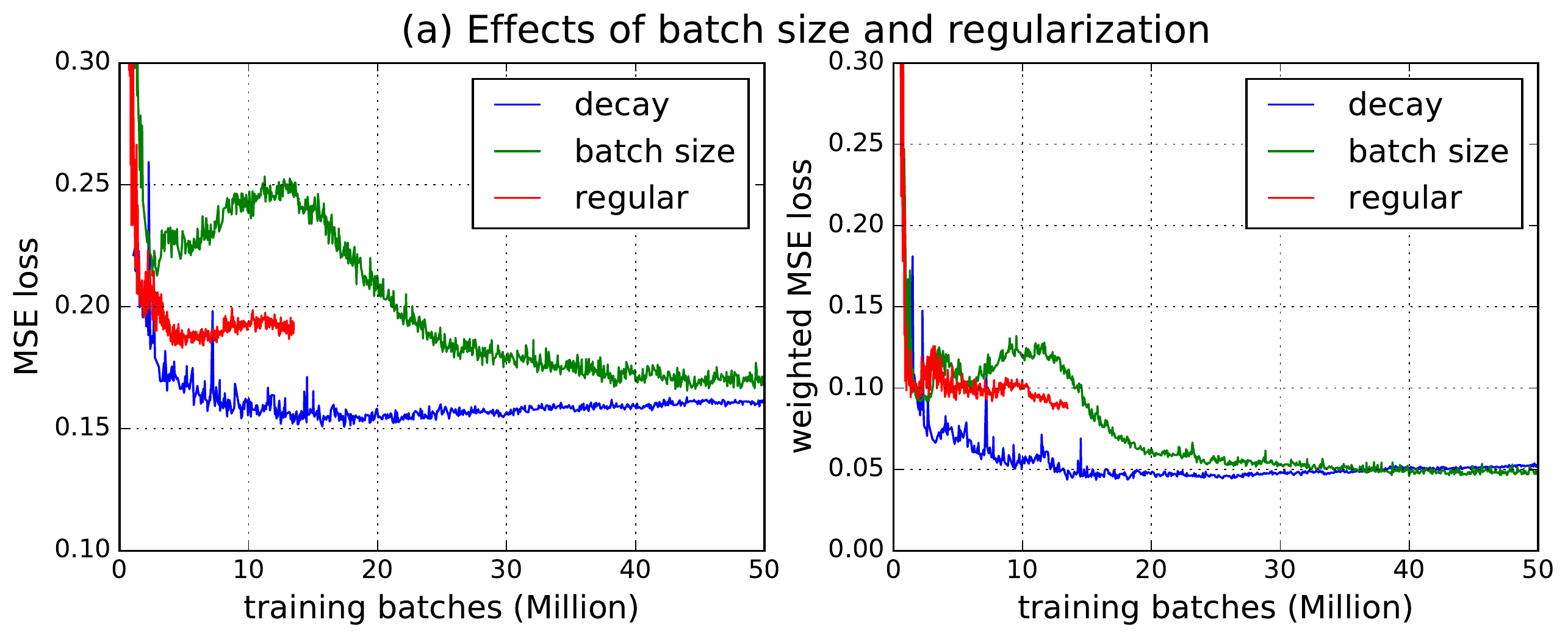}
		\caption{Comparison of the convergence speed of training by using different hyperparameters. The hyperparameter description is shown in Table~\ref{Tb:OfflineExp3}.}
		\label{fig:OfflineExp3}
	\end{figure}

	We also evaluate the performance improvement brought by the dueling architecture in Fig.~\ref{fig:OfflineExp1}. Comparing the results of using the dueling architecture (annotated by `dueling' in the figure) and not (`without dueling'), it can be seen that, the dueling architecture improve the convergence speed dramatically. The intuition behind the results is the dueling architecture could help remove the reward variance of the same action under different states by $V(s)$, and guide the action-value network $A(s, a)$ to focus more on differentiating between different actions. As a result, the policy network learning is accelerated. In Fig.~\ref{fig:OfflineExp3}, we evaluate the influences of different super parameters for training the model, including the learning rate, regularization penalties and batch sizes. The parameter setup is listed in Table~\ref{Tb:OfflineExp3}. As can be observed, lower learning rate (`base' and `learning rate') makes the learning method converge more smoothly and larger batch size (`base' and `batch size') and lower regularization penalty (see `base' and `regular') make the learning method converge more closely to the optimal solution. This is because there is strong variance in the rewards. Lower learning rate and larger batch size helps to reduce the variance in the batched training data, and lower regularization penalty allows a larger searching space for variables.
	
	\begin{table}[tbp]
		\centering
		\caption{Hyperparameter setup of Fig.~\ref{fig:OfflineExp3}.}
		\label{Tb:OfflineExp3}
		\begin{tabular}{lccc}
			\hline
			ID & Learning rate & Regularization & Batch size\\
			\hline
			base & 1.0e-5 & 1.0e-5 & 50k \\
			learning rate & 1.0e-4 & 1.0e-5 & 50k\\
			batch size & 1.0e-5 & 1.0e-5 & 10k \\
			regular & 1.0e-5 & 1.0e-3 & 50k \\
			\hline
		\end{tabular}
	\end{table}
		
	\subsection{Online Serving Experiments}
	
	In this section, we present the experimental results of conducting a bucket experiment on a popular sponsored search platform. The sponsored search platform charge the advertisers for the click on their advertisements according to GSP auction mechanism. We split a small percentage (about 2\%) of traffic from the whole online traffic by hashing the user IDs, IP addresses, etc., and deploy the learned ranking function online for advertisement selection and pricing. The following business metric are measured to see the improvement brought by the proposed method. (1) Revenue-Per-Mille (RPM): the revenue generated per thousand impressions; (2) Price-Per-Click (PPC): the average price per-click determined by the auction; (3) Click-through-rate (CTR). We employ the three business metrics because we are optimizing towards the platform revenue and user experience as in Eq.~(\ref{eq:rewardexp}). RPM is determined by the product of CTR and PPC. From the change of CTR and PPC, we can also induce the improvements of the advertisers' saling efficiency, i.e. increase in CTR and decrease in PPC means the advertisers can attract more customers by less spending on advertisement serving.
	
	In this work, a new ranking function is proposed by adding the terms reflecting the user engagement and the advertisers' gain, and a reinforcement learning method is presented for learning the optimal parameter of this term. In the online experiments, we want to verify the performance improvement brought by the new ranking function and the improvement from the reinforcement learning method. To evaluate the effectiveness of the new ranking function, we compare the proposed ranking function with the one proposed by Lahaie and McAfee \cite{lahaie2011efficient}. The ranking function in \cite{lahaie2011efficient} has been employed by many companies and proved to be efficient in online advertising auctions. In the experiment, we use this ranking function as the baseline and set its parameter (the exponential term) by brute force searching in the same manner as Section~\ref{sub:OfflineSimExp}. For the proposed method, we use both the ranking functions learned by the brute force method in Section~\ref{sub:OfflineSimExp} and the ranking function learned by the reinforcement learning method. The comparison results on two continuous days' data are shown in Table~\ref{Tb:OnlineExp1}. As is observed, compared to the method~\cite{lahaie2011efficient}, our ranking function with parameters searched by brute force method is capable of devlivering 2.5\% of RPM growth with 1.1\% of CTR increase and 1.4\% of PPC increase. This observation confirms the effectiveness of the proposed ranking function in improving the platform's performance. For the ranking function learned by reinforcement learning method, we observe a 2.5\% RPM growth which is arisen majorly from the CTR increase (2.0\% of CTR gain). We interpret the difference between the reinforcement learning method and 'brute force' method by the fact that the reinforcement learning method dose not converge to the exact value of brute force method, and it use a more elaborate set of features for learning. From the business side, we can find that the new ranking function can improve more user engagement by attracting more user clickness on the presented advertisements. For the platform, the increase of RPM brings in more efficiency in earn platform profit, and for the advertisers, the RPM growth is driven by the CTR increase while there is a little increase in PPC (for the learned ranking function), this means, the advertisers only need pay a litter more money to attract more potential buyers.
	
\begin{figure}
	\includegraphics[width=0.46\textwidth]{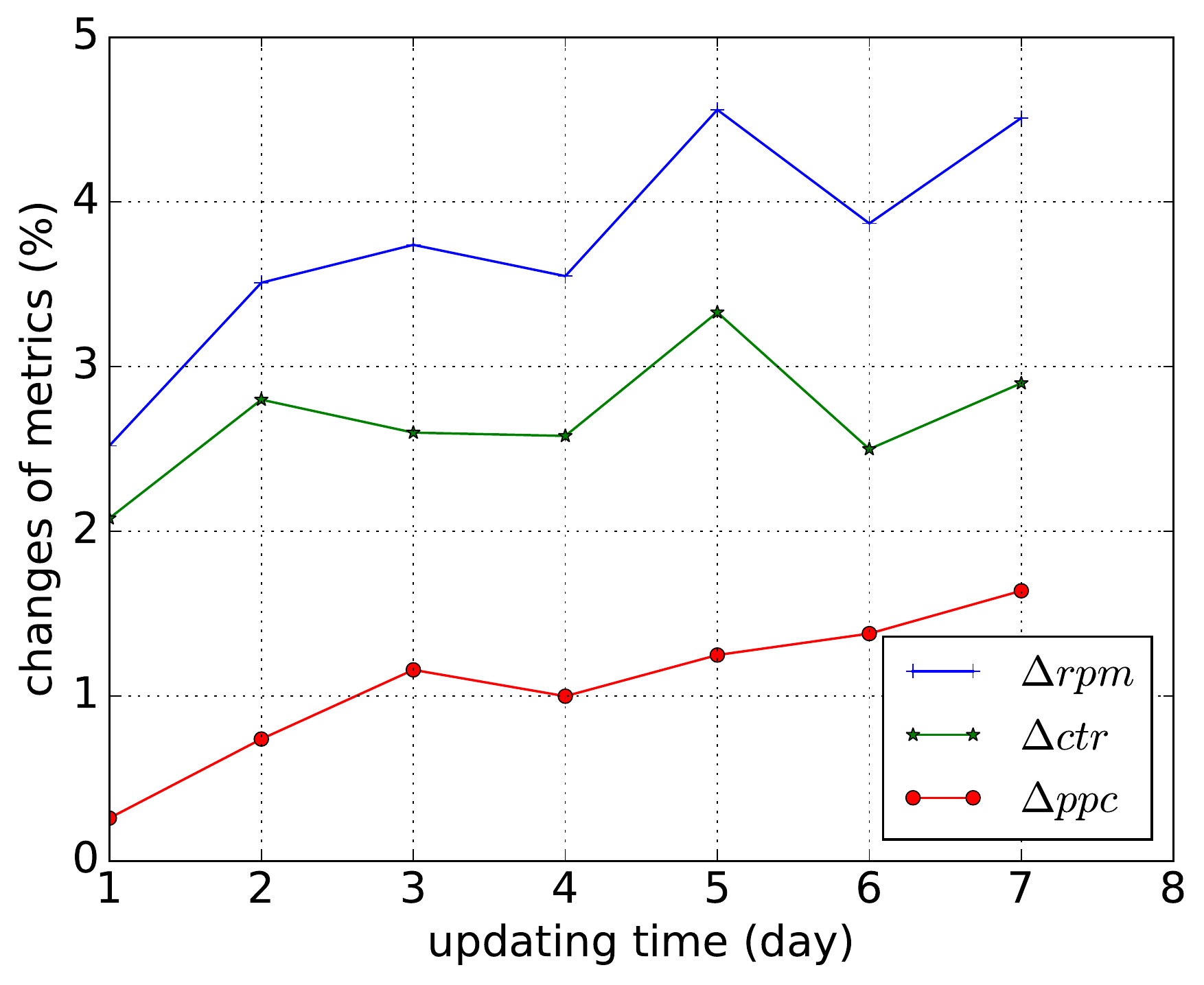}
	\caption{Experimental results illustrating the business metric changes during the online update of the proposed method in Section~\ref{Sec:OnlineLearning}.}
	\label{fig:OnlineExp}
	\end{figure}

	Section~\ref{Sec:OnlineLearning} introduces the online evolution strategy method for tuning the policy networks based on online data. In this experiment, we evaluate the online performance changes in seven continuous days to confirm the performance increases brought by online learning method. To disturb strategy actions in $\pi_{\theta_\pi}(s)$, we add a gaussian noise $G(0, \delta^2)$ with mean 0 and variance $\delta^2=0.01$ to the parameters ${\theta_\pi}$ of $\pi_{\theta_\pi}(s)$. We split the traffic of the test bucket into a number of splits and apply each perturbed policy networks to one of them. The buckets and the user feedback history are collected from the data logs to compute the updates in Eq.~(\ref{eq:onlineupdate}). The average performance of the test bucket are shown in Fig.~\ref{fig:OnlineExp}. As can be seen, all the three business metrics improves during the days. Compared with the baseline ranking function \cite{lahaie2011efficient} (whose parameter stay unchanged during the days), the RPM grows from 2\% to 4\%, CTR grows from 2\% to about 3\%. The results indicate the effectiveness of the online updating method. We also find the PPC grows because there is no constraints added on it. In the future work, we will try to add constraints to limit PPC increase to generate more return for advertisers.
	
	\begin{table}[tbp]
	\centering
	\caption{Experimental results comparing the performance of the ranking function in \cite{lahaie2011efficient}, the brute force searched ranking function in Section~\ref{sub:OfflineSimExp} and the one learned by reinforcement learning method in Section~\ref{sub:DdpgOffline}.}
	\label{Tb:OnlineExp1}
	\begin{tabular}{lccc}
	\hline
	Metrics (\%) & $\Delta_{rpm}$ & $\Delta_{ctr}$ & $\Delta_{ppc}$\\
	\hline
	McAfee~\cite{lahaie2011efficient} & 0.00 & 0.00 & 0.00 \\
	brute force& 2.55 & 1.12 & 1.45\\
	offline learning & 2.52 & 2.08 & 0.26 \\
	\hline
	\end{tabular}
	\end{table}
	
	\section{Conclusions}
	
	It is commonly accepted when building a commercial sponsored search platform, besides the intermediate platform revenue, the users' engagement and the advertisers' return 
	are also important to the long-term profit of the commercial platform. In this work, we design a new ranking function by incorporating these factors together. However,
	these additional terms increase the complexity of the ranking function. In this work, we propose a reinforcement learning framework to optimize the ranking function 
	towards the optimal long-term profit of the platform. As is known, the reinforcement learning works in a trail-and-error manner. To allow
	adequate exploration without hurting the performance of the commercial platform, we propose to initialize the ranking function by an offline learning procedure conducted in a simulated
	sponsored search environment, followed by an online learning module which updates the model adaptively to online data distribution. Experimental results confirms the effectiveness of the
	proposed method.
	
	
	In the future, we will focus on the following directions: (1) Sequential user behavior simulation. The environment simulation method introduced in Section~\ref{subsec:replay} is
	limited to one time advertisement serving without considering the correlation between sequential user behaviors. We plan to try generative models like GAN \cite{Goodfellow14}
	to model the continuous user behaviors. (2) The proposed method has the ability to improve advertisers' return-per-cost by adding the purchase amount into the reward term.
	Due to traffic limitation, we did not investigate on this effect, in future work, we will increase the online testing traffic to measure the gain brought by adding the purchase amount reward.
	
	
	
	
	
	\bibliographystyle{ACM-Reference-Format}
	\bibliography{rl-ad-rank} 
	
\end{document}